\documentstyle[epsfig,portland]{aipproc}
\begin{document}
\title{Self-force approach for radiation reaction}
\author{Lior M. Burko}
\address{Theoretical Astrophysics, California Institute of Technology,
Pasadena, California 91125}
\maketitle

\begin{abstract}
We overview the recently proposed mode-sum regularization prescription
(MSRP) for the
calculation of the local radiation-reaction forces, which are crucial for
the orbital evolution of binaries. We then describe some new results which 
were obtained using MSRP, and discuss their importance for
gravitational-wave astronomy.
\end{abstract}

The problem of including the radiation-reaction (RR) forces in the orbital
evolution of a binary is a long-standing open problem. This problem is as
yet unresolved even in the extreme mass-ratio limit, with the particle
orbiting a non-rotating black hole, although there has been a remarkable
progress obtained from various directions \cite{schutz}. 
The conventional approach is to consider the fields in the far zone, and
then use a balance
argument to relate the far-zone fields to the local properties of the
particle. The generic failure of such approaches \cite{hughes} prompted
the idea to calculate the {\it local} forces acting on the particle,
including
the RR forces. In the following we discuss the RR forces acting on a
scalar point-like charge, but for electric or gravitational charges the
basic ideas are similar. The RR force ${^{\rm RR}}F^{\mu}$ which acts on a 
point-like scalar charge $q$ is given by \cite{quinn-wiseman}
\begin{equation}
{^{\rm
RR}}F^{\mu}(\tau)=q^2\left[\frac{1}{3}\left(\ddot{u}^{\mu}-u^{\mu}\dot{u}_{\nu}
\dot{u}^{\nu}\right)+\frac{1}{6}{\cal
R}^{\mu}+\int_{-\infty}^{\tau}\nabla^{\mu}G_{\rm R}\,d\tau'\right],
\label{rr-force}
\end{equation}
where ${\cal R}^{\mu}=R^{\mu}_{\;\nu}u^{\nu}+
R_{\nu\sigma}u^{\nu}u^{\sigma}u^{\mu}-\frac{1}{2}R^{\nu}_{\;\nu}u^{\mu}$,  
$R_{\nu\sigma}$ is the Ricci tensor, $u^{\mu}$ is the charge's 
4-velocity, a dot denotes (covariant) derivative with respect to proper
time $\tau$, $\nabla_{\mu}$ denotes covariant differentiation, and $G_{\rm
R}$ is the retarded Green's function. The first term is a local 
Abraham-Lorentz-Dirac type damping force, the second is a local force,
which couples to Ricci-curvature and preserves conformal invariance, and
the third is the so-called ``tail''
term, which arises from the failure of the Huygens principle in curved
spacetime. The greatest problems in the calculation of the RR forces lurk  
in the tail term, because it requires the knowledge of $G_{\rm R}$ along
the entire past world line of the charge. In addition, the self field of
any particle diverges at the position of the particle, and the calculation
of the RR forces will have to handle the infinities connected with the
self field by providing a regularization prescription.

Recently, Ori proposed to approach the RR problem via mode decomposition 
\cite{ori-rr}. Ori observed, that the individual Fourier-harmonic modes of
the self field are bounded, also for a point-like particle, although the
sum over all modes diverges. This observation is very useful, because the
calculation of the individual modes is relatively easy. 
This still leaves the second, harder problem of having a regularization
prescription to handle the mode sum. Very recently, Ori suggested MSRP 
\cite{ori-unpublished}, which is very successful for the few simple cases
to which it has already been applied. In what follows, we overview MSRP
very briefly, and describe some of the recent results which were obtained
using it.

The tail part of the RR force 
can be decomposed into stationary Teukolsky modes, and
then summed over the frequencies $\omega$ and the azymuthal numbers $m$.
This force equals then the limit $\epsilon\to 0^-$ of the
sum over all $\ell$ modes, of the difference between the force sourced by
the entire world line (the bare force ${^{\rm bare}}F_{\mu}^{\ell}$) and
the force sourced by the half-infinite world
line to the future of $\epsilon$, where the particle has proper time
$\tau=0$, and $\epsilon$ is an event along the past ($\tau<0$) world
line. Next, we seek a function $h^{\ell}_{\mu}$ which is
independent of $\epsilon$, such that the series $\sum_{\ell}({^{\rm
bare}}F_{\mu}^{\ell}-h^{\ell}_{\mu})$ converges. Once such a function is
found, the regularized self force is then given by ${^{\rm 
tail}}F_{\mu}=\sum_{\ell}({^{\rm bare}}F_{\mu}^{\ell}-h^{\ell}_{\mu})-
d_{\mu}$, where $d_{\mu}$ is a finite valued function. MSRP
\cite{ori-unpublished} then shows, from a local integration of 
$G_{\rm R}$, that 
$h^{\ell}_{\mu}=a_{\mu}\ell+b_{\mu}+c_{\mu}\ell^{-1}$. MSRP also
provides an algorithm for the calculation of the
functions $a_{\mu}, b_{\mu}, c_{\mu}$ and $d_{\mu}$ analytically. It has 
been conjectured, that for all orbits $a_{\mu}=0=c_{\mu}$. (It was found
to be true for all the special cases calculated so far.) If this is
indeed the case, the tail force is given by 
\begin{equation}
{^{\rm tail}}F_{\mu}=
\sum_{\ell}({^{\rm bare}}F_{\mu}^{\ell}-b_{\mu})-d_{\mu}. 
\label{tail-force}
\end{equation}
Note that $b_{\mu}$ is just the limit ${^{\rm
bare}}F_{\mu}^{\ell\to\infty}$, and that 
${^{\rm bare}}F_{\mu}^{\ell}$ can
be computed using the Teukolsky formalism. Alternatively, $b_{\mu}$ can
also be calculated analytically using MSRP. The only remaining problem
then, is to calculate $d_{\mu}$. Even though there is an algorithmic way
to calculate $d_{\mu}$, this calculation is by no means easy.  
Although MSRP has been developed as yet only for very simplified cases, 
the approach is likely to be susceptible of generalization also for more 
realistic cases. If robust, MSRP can be of the greatest importance for the
calculation of templates for gravitational-wave detection.

Table \ref{table1} displays the values of the MSRP parameters for the
cases which have already been calculated. Note that the motion is not
necessarily geodesic. The data in Table \ref{table1} may suggest the
conjecture that $d_{\mu}$ exactly equals the sum of the two local terms of
Eq. (\ref{rr-force}) (or its analog for other charge types). If this
hypothesis is proved to hold in general, then a truely remarkable thing
happens: the full RR force can be calculated directly from Eq.
(\ref{tail-force}) when $d_{\mu}$ is ignored. It should be emphasized that
presently the support for this far-reaching hypothesis is only the special
cases listed in Table \ref{table1}. 

\begin{table}
\caption{Values of the MSRP parameters. All spacetimes are
spherically symmetric. For all the
cases $a_{\mu}=0=c_{\mu}$. The charge's spin is $s$. A $*$
denotes the cases for which $b_{\mu}$ and $d_{\mu}$ were inferred
indirectly, and a $\dagger$ denotes cases where the values of $b_{\mu}$
were corroborated numerically.  
Here, $I_a={{_2}F}_1(1/2,1/2;1;v^2)$ and $I_b={{_2}F}_1(1/2,3/2;1;v^2)$,
$v^2=-(\,d\varphi/\,dt)^2r^2/g_{tt}$, $a^{\mu}$ is the 4-acceleration, 
$a^2=\dot{u}_{\alpha}\dot{u}^{\alpha}$, and $k_{\mu}=
\frac{q^2}{3}(\ddot{u}_{\mu}-u_{\mu}a^2)$.}
\label{table1}
\begin{tabular}{lllccc}
Type of orbit & Spacetime & $s$ & $\mu$ & $b_{\mu}/[-q^2/(2r^2)]$ &
$d_{\mu}$ \\ 
\tableline
Static \cite{ori-unpublished,burko-unpublished}  & Minkowski & $0,1$ & $t$
& $0$ & $0$
\\
Static \cite{ori-unpublished,burko-unpublished}  & Minkowski & $0,1$ & $r$
& $1$ & $0$
\\
Static$^\dagger$ \cite{ori-unpublished,burko-cqg}  & Schwarzschild & $0$ &
$r$ &
$(r-M)/(r-2M)$ & $0$ \\
Circular$^\dagger$ \cite{ori-unpublished,burko-unpublished}  & Minkowski &
$0$ & $r$ &
$(2I_a-I_b)/u^t$ & $0$\\
Circular$^{*\dagger}$ \cite{ori-unpublished,burko-unpublished}  &
Minkowski & $0$ & $t$
& $0$ &
$k_t$\\
Circular$^{*\dagger}$ \cite{ajp}  & Minkowski & $1$ & $t$ & $0$ &
$2k_t$\\
Circular$^\dagger$ \cite{ori-unpublished,burko-prl}  & Schwarzschild & $0$
& $r$ & 
$(2I_a-\frac{r-3M}{r-2M}I_b)/(u^t\sqrt{-g_{tt}})$ & $0$\\
Radial \cite{barack-ori}  & Minkowski & $0$ & $r$ &
$1-\dot{r}^2+r\ddot{r}$ 
& $k_r$ \\
Radial \cite{barack-ori}  & $g_{tt}(r)g_{rr}(r)=-1$ & $0$ & $r$ &
$1-\dot{r}u_r+ra_r$ & 
$k_r+q^2{\cal R}_{r}/6$ \\
Radial \cite{barack-ori}  & $g_{tt}(r)g_{rr}(r)=-1$ & $0$ & $t$ &
$-\dot{r}u_t+ra_t$ &
$k_t+q^2{\cal R}_{t}/6$ \\
Static$^*$ \cite{burko-unpublished} & Reissner-Nordstr\"{o}m & $1$ & $t$ & 
$0$ & $q^2{\cal R}_{t}/3$ 
\end{tabular}
\end{table}   

Next, we present some results which were obtained by application of this
new approach. For the case of a static, minimally-coupled, massless scalar
charge in Schwarzschild, the self force is known to equal 
zero \cite{static-scalar}. 
For a static electric charge $q$ in Schwarzschild the
self force is  known to be purely radial and to be given by 
$f_r=q^2Mr^{-3}(1-2M/r)^{-1/2}$ \cite{smith-will}. These results 
were recovered using MSRP in \cite{burko-cqg}. Note that for these two
simple static cases the solution for the modes can be obtained
analytically. In general, however, this is not expected to be possible,
and the solution can be obtained only numerically. 

\begin{figure}
\centerline{\epsfig{file=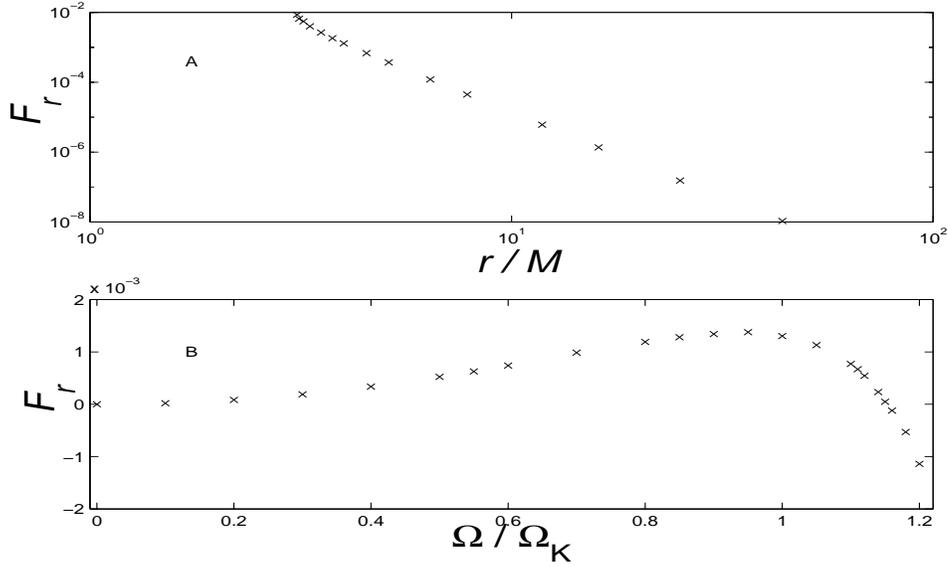,height=3.0in,width=5.0in}}
\vspace{10pt}
\caption{The radial RR force acting on a scalar charge in uniform
circular orbit around a Schwarzschild black hole. Top panel (A): $F_r$ as
a function of $r/M$ for geodesic orbits.
Bottom panel (B): $F_r$ as a function of the angular velocity $\Omega$ in
units of the Keplerian angular velocity $\Omega_K$, when the orbit is at
$r=4M$.} 
\label{fig1}
\end{figure}

The case of a scalar charge $q$ in uniform circular
orbit around a Schwarzschild black hole was recently considered in
\cite{burko-prl}. The RR force was calculated numerically without any
simplifying assumptions, such as far field or slow motion, and the
solution
is fully relativistic. Both the temporal and the azimuthal, dissipative
components and the radial, conservative component of the RR force were
computed. Figure \ref{fig1} displays the behavior of the radial component
of the RR force for both geodesic and non-geodesic orbits. In the slow
motion and far field limits the force is repulsive, and behaves like 
$ 
{^{\rm RR}}F_r\approx q^2M^2\Omega^2/r^2.
$ 
However, in strong fields the force grows faster, and for fast motion is
changed from repulsive to attractive. This expression for the radial,
conservative RR force may be very important for the detection of 
gravitational waves, and also for gravitational-waves astronomy. The
conservative radial force causes an additional precession of the
periastron of the particle's orbit, and thus induces a change in the
frequency and phase of the emitted radiation \cite{burko-br}. Although
the radial self force has been obtained only for the simple case of a
point-like scalar charge, the result indicates that one can expect a
non-zero periastron precession also for a small mass. However, the
magnitude of the effect will very reasonably depend on the type of the
charge. A large-magnitude effect can cause the entire search algorithm
to fail in the very detection of the signal (depending also on the size of  
the template library), and a small effect will introduce
errors in the parameters of the observed binary, namely, the wave form
would fit the template of a system with parameters different from the
parameter of the actual binary. Note that the conservative
force depends not only on the radiative modes of the field, but also on
the non-radiative modes \cite{burko-br}. 

I thank Leor Barack and Amos Ori for discussions and for letting me use
their results before their publication. 
This work was supported by NSF grants AST-9731698 and PHY-9900776 and by
NASA grant NAG5-6840.


\begin{references}
\bibitem{schutz} See, e.g., B. F. Schutz, Class. Quantum Grav. {\bf 16},
A131 (1999).
\bibitem{hughes} S. A. Hughes, these proceedings; gr-qc/9910091.
\bibitem{quinn-wiseman} T. C. Quinn and A. G. Wiseman, in preparation.
\bibitem{ori-rr} A. Ori, Phys. Lett. {\bf A202}, 347 (1995); Phys. Rev. 
D{\bf 55}, 3444 (1997).
\bibitem{ori-unpublished} A. Ori, unpublished; For a brief
outline of the proposed regularization method, see L. Barack and A. Ori,
gr-qc/9911040.
\bibitem{burko-unpublished} L. M. Burko, unpublished.
\bibitem{burko-cqg} L. M. Burko, Class. Quantum Grav. {\bf 17}, 227
(2000).
\bibitem{ajp} L. M. Burko, Am. J. Phys. (in press), also gr-qc/9902079.
\bibitem{burko-prl} L. M. Burko, in preparation.
\bibitem{barack-ori} L. Barack and A. Ori, in preparation. 
\bibitem{static-scalar} A. I. Zel'nikov and V. P. Frolov, Sov. Phys. JETP
{\bf 55}, 191 (1982); A. G. Wiseman, in preparation; A. E. Mayo, Phys.
Rev. D {\bf 60}, 104044 (1999).
\bibitem{smith-will} A. G. Smith and C. M. Will, Phys. Rev. D {\bf 22},
1276 (1980).
\bibitem{burko-br} L. M. Burko, in preparation. 
\end{references}
\end{document}